\documentstyle[12pt,aaspp4]{article}
\begin{document}
\newcommand{\up}[1]{\ifmmode^{\rm #1}\else$^{\rm #1}$\fi}
\newcommand{\zdot}{\makebox[0pt][l]{.}}
\newcommand{\upd}{\up{d}}
\newcommand{\uph}{\up{h}}
\newcommand{\upm}{\up{m}}
\newcommand{\ups}{\up{s}}
\newcommand{\arcd}{\ifmmode^{\circ}\else$^{\circ}$\fi}
\newcommand{\arcm}{\ifmmode{'}\else$'$\fi}
\newcommand{\arcs}{\ifmmode{''}\else$''$\fi}

\title{The Araucaria Project. Near-Infrared Photometry of Cepheid Variables
in the Sculptor Galaxy NGC 55
\footnote{Based on observations obtained with the ESO VLT for 
Large Programme 171.D-0004}
}
\author{Wolfgang Gieren}
\affil{Universidad de Concepci{\'o}n, Departamento de Fisica, Astronomy Group,
Casilla 160-C, Concepci{\'o}n, Chile}
\authoremail{wgieren@astro-udec.cl}
\author{Grzegorz Pietrzy{\'n}ski}
\affil{Universidad de Concepci{\'o}n, Departamento de Fisica, Astronomy
Group,
Casilla 160-C,
Concepci{\'o}n, Chile}
\affil{Warsaw University Observatory, Al. Ujazdowskie 4, 00-478, Warsaw,
Poland}
\authoremail{pietrzyn@hubble.cfm.udec.cl}
\author{Igor Soszy{\'n}ski}
\affil{Universidad de Concepci{\'o}n, Departamento de Fisica, Astronomy Group, 
Casilla 160-C, Concepci{\'o}n, Chile}
\affil{Warsaw University Observatory, Al. Ujazdowskie 4, 00-478, Warsaw,
Poland}
\authoremail{soszynsk@astro-udec.cl}
\author{Fabio Bresolin}
\affil{Institute for Astronomy, University of Hawaii at Manoa, 2680 Woodlawn 
Drive, 
Honolulu HI 96822, USA}
\authoremail{bresolin@ifa.hawaii.edu}
\author{Rolf-Peter Kudritzki}
\affil{Institute for Astronomy, University of Hawaii at Manoa, 2680 Woodlawn 
Drive,
Honolulu HI 96822, USA}
\authoremail{kud@ifa.hawaii.edu}
\author{Jesper Storm}
\affil{Astrophysikalisches Institut Potsdam, An der Sternwarte 16, D-14482
Potsdam, Germany}
\authoremail{jstorm@aip.de}
\author{Dante Minniti}
\affil{Departamento de Astronomia y Astrofisica, Pontificia Universidad Cat{\'o}lica
de Chile, Casilla 306, Santiago 22, Chile}
\authoremail{dante@astro.puc.cl}

\begin{abstract}
We have obtained deep images in the near-infrared J and K filters of four fields in the
Sculptor Group spiral galaxy NGC 55 with the ESO VLT and ISAAC camera. For 40 long-period
Cepheid variables in these fields which were recently discovered by Pietrzy{\'n}ski et al., 
we have determined mean J and K magnitudes from observations at two epochs,
and derived distance moduli from the observed
PL relations in these bands. Using these values together with the previously measured distance moduli
in the optical V and I bands,  we have determined a total mean reddening of the NGC 55 Cepheids of
E(B-V)=0.127 $\pm$ 0.019 mag, which is mostly produced inside NGC 55 itself.
For the true distance modulus of the galaxy, our multiwavelength analysis yields
a value of 26.434 $\pm$ 0.037 mag (random error), 
corresponding to a distance of 1.94 $\pm$ 0.03 Mpc. This value is tied to an adopted 
true LMC distance modulus of 18.50 mag. The systematic uncertainty of our derived
Cepheid distance to NGC 55
(apart from the uncertainty on the adopted LMC distance) is $\pm$4\%, with the main 
contribution likely to come from the effect of blending of some of the Cepheids with
unresolved companion stars. The distance of NGC 55 derived from our multiwavelength Cepheid
analysis agrees within the errors with the distance of NGC 300, strengthening the case
for a physical association of these two Sculptor Group galaxies.
\end{abstract}


\section{Introduction}

The effectiveness of using multiwavelength optical and near-infrared (NIR) observations
of Cepheid variables for distance determination of galaxies has been known for a long
time (McGonegal et al. 1982; Madore \& Freedman 1991). Only recently however the
technical problems with obtaining accurate and reliable NIR photometry for faint objects
in dense regions have been solved. Using NIR observations of Cepheids provides a number
of advantages for accurate distance work. First, the total and differential reddening
is significantly reduced in comparison to the optical bandpasses. Second, the Cepheid
PL relation becomes steeper toward longer wavelengths, and its intrinsic dispersion
becomes smaller, both factors helping in deriving a more accurate distance. Third, 
metallicity effects on the PL relation in the near-IR are expected to be less important than at
optical wavelengths (Bono et al. 1999). Fourth, and very importantly from an observational
point of view, the amplitudes of variability are significantly smaller in the NIR
than at optical wavelengths, so even random single-epoch observations
approximate the mean magnitude reasonably well. If the period and optical light curve
of a Cepheid is accurately known, it is possible to derive its mean magnitude in the NIR bands
with an impressive 1-2\% accuracy from just one single-epoch observation (Soszy{\'n}ski
et al. 2005). 

Simultaneously studying NIR and optical PL relations of Cepheids provides another
important advantage. By combining the observed distance moduli in the optical
and NIR it is possible to derive both the total reddening and the true distance modulus
with very high accuracy. This has been demonstrated in the previous papers of
this series which reported the distances to NGC 300 (Gieren et al. 2005a), IC 1613
(Pietrzy{\'n}ski et al. 2006a), NGC 6822 (Gieren et al. 2006) and NGC 3109 (Soszy{\'n}ski
et al. 2006) derived by this method. For all these galaxies, we were able to determine
their distances with respect to the LMC with a $\sim$3\% accuracy from our technique. The
multiband closure on the total reddening estimates, through the use of the Wesenheit function
and near-infrared data, has made it possible to achieve these accuracies.

NGC 55 is a highly inclined late-type galaxy classified as SB(s)m
in the NASA/IPAC Extragalactic Database. It bears some resemblance to the LMC and is one of the
approximately 30 known members of the Sculptor Group (Jerjen et al. 2000).
NGC 55 was included in the list of target galaxies of our ongoing
Araucaria Project (Gieren et al. 2005b) because of its relative proximity which allows accurate
photometry and spectroscopy of the different stellar distance indicators scrutinized
in our project, and because existing oxygen abundance determinations from H II regions indicate a low
metallicity of about 0.25 solar, close to the SMC (e.g. Lee et al. 2006, and references therein) which makes NGC 55
the lowest-metallicity spiral galaxy in our sample. This is important because one of our
main goals in the Araucaria Project is to determine the effect of metallicity on the various stellar methods
of distance determination we are investigating. Also, color images of NGC 55 clearly suggested
the existence of an abundant young stellar population in this galaxy, suggesting the presence
of abundant blue supergiant and Cepheid populations, both types of objects being very useful
for distance determination. In fact, we have discovered more than 100 blue supergiants in
NGC 55 and have obtained low-resolution spectra for abundance analysis with VLT/FORS. The abundance
results we will obtain from these data will give us an opportunity of independently check
on the nebular oxygen abundances, and will allow us to determine the distance of the galaxy
with the Flux-Weighted Gravity-Luminosity Relation introduced by Kudritzki et al. (2003).

There are several distance determinations to NGC 55 based on different methods in
the literature, which have been collected in Table 4. They will be briefly discussed
in section 4. These previous distance estimates
have yielded results which range from 1.34 Mpc (Pritchet et al. 1987) to 2.30 Mpc (Van de Steene
et al. 2006). This considerable discrepancy hints at relatively large errors
in one or several of these previous determinations, and a more accurate distance determination
was therefore clearly desirable. Since no previous surveys for Cepheid variables
had been conducted in NGC 55, we have carried out such a wide-field Cepheid survey
in optical V and I filters which resulted in the discovery of 143 Cepheids
(Pietrzy{\'n}ski et al. 2006b; hereafter Paper I). 
From 130 Cepheids with periods longer than 10 days, we constructed PL relations
in V, I and the reddening-independent (V-I) Wesenheit band, yielding as a best estimate
of the distance to NGC 55 a value of 1.91 $\pm$ 0.10 Mpc.In this paper, we extend the
light curve coverage for 40 Cepheids in NGC 55 to the NIR J and K bands. We then utilize the
multiband VIJK data for these stars for an accurate determination of the total (average)
interstellar extinction to the Cepheids in NGC 55, and determine an improved distance
to the galaxy. As mentioned above, the extension of the Cepheid work to the NIR is a fundamental step to reduce
the systematic error in the Cepheid distance, mostly by decreasing the sensitivity of the result
to reddening and effects of metallicity on the period-luminosity relation.

The paper is composed as follows. In section 2 we describe the NIR observations, data reductions
and calibration of our photometry. In section 3 we derive the J- and K-band Cepheid PL relations
in NGC 55 from our data and determine the true distance modulus to NGC 55 from a multiwavelength
analysis. In section 4, we discuss our results, and in section 5 we summarize our conclusions.

\section{Observations, Data Reduction and Calibration}

We used deep J- and K-band images recorded with the 8.2 m ESO Very Large Telescope equipped
with the Infrared Spectrometer and Array Camera (ISAAC). Figure 1 shows the location of the four
2.5 x 2.5 arcmin fields observed in service mode on 7 nights between 18-07-2004 and 21-09-2004. 
The coordinates of the field centers were chosen in such a way as to maximize the number
of Cepheid variables observed and optimize their period distribution. Each field was
observed in both NIR bands two times, on two different nights, and therefore at different
pulsation phases of the Cepheids in these fields. The observations were carried out
using a dithering technique, with a dithering of the frames following a random pattern
characterized by typical offsets of 15 arcsec. The final frames in the J and K bands
were obtained as a co-addition of 24 and 164 single exposures obtained with integration times
of 30 and 15 s, respectively. Thus, the total exposure time for a given observation was
12 minutes in J and 41 minutes in K. The observations were obtained under excellent seeing conditions,
typically around 0.5 arcsec. Standard stars on the UKIRT system (Hawarden et al. 2001) were
observed along with the science exposures to allow an accurate transformation of the instrumental
magnitudes to the standard system.

The images were reduced using the program JITTER from the ECLIPSE package developed by ESO to reduce
near-IR data. The point-spread function (PSF) photometry was carried out with the DAOPHOT and
ALLSTAR programs. The PSF model was derived iteratively from 20 to 30 isolated bright stars following
the procedure described by Pietrzy{\'n}ski et al. (2002). In order to convert our profile
photometry to the aperture system, aperture corrections were computed using the same stars
as those used for the calculation of the PSF model. The median of the aperture corrections obtained
for all these stars was finally adopted as the aperture correction for a given frame. The aperture
photometry for the standard stars was performed with DAOPHOT using the same aperture as the one
adopted for the calculation of the aperture corrections. 

The astrometric solution for the observed fields was performed by cross-identification of the
brightest stars in each field with the Infrared Digitized Sky Survey 2 (DSS2-infrared) images.
We used programs developed by Udalski et al. (1998) to calculate the transformations between 
the pixel grid of our images and equatorial coordinates of the DSS astrometric system. The internal
error of the transformation is less than 0.3 arcsec, but systematic errors of the DSS coordinates
can be up to about 0.7 arcsec. 

In order to perform an external check on our photometric zero points, we tried to compare the
magnitudes of stars in the 2MASS Point Source Catalog located in our NGC 55 fields with
our own photometry. Unfortunately, even the brightest stars in our dataset whose photometry is
not affected by nonlinearity problems are still very close to the faint magnitude limit of
the 2MASS catalog and have 2MASS magnitudes with formal errors of $\sim$0.2 mag. Moreover,
all our NGC 55 fields are located in regions of high stellar density (see Figure 1), and most
of the 2MASS stars turn out to be severely blended as seen at the higher resolution of our VLT/ISAAC images.
It was therefore not possible to carry out a reliable comparison of the two photometries. However,
since our reduction and calibration procedure is extremely stable, we should have achieved the same
typical zero point accuracy we were able to achieve in the previous near-infrared studies
of Local Group galaxies where a comparison with 2MASS magnitudes for a common set of stars
was possible. From this argument, we expect that our photometric zero points are determined 
to better than $\pm$0.03 mag in both J and K filters.

Our four fields in NGC 55 contain a subset of 40 of the 143 Cepheids reported in Paper I. All
individual observations in J and K we obtained are presented in Table 1 which lists the star's IDs,
heliocentric Julian day of the observations, and the measurements in J and K with their
standard deviations. For most of the Cepheids we collected two observations per given filter.
For some objects we obtained only one observation in the J or K band as a consequence
of the locations of these objects close to the edge of the observed field.

\section{Near-Infrared Period-Luminosity Relations}

All individual J and K measurements reported in Table 1 were transformed to the mean magnitudes 
of the Cepheids using the recipe given by Soszy{\'n}ski et al. (2005). The corrections to derive
the mean magnitudes from the observed random-phase magnitudes were calculated by taking advantage of
the complete V- and I-band light curves from Paper I, exactly in the way
as described in the Soszy{\'n}ski et al. paper.
For the vast majority of the Cepheids,
the mean J and K magnitudes determined from the two individual random-phase magnitudes agree
very well. We were helped by the fact that the optical observations in Paper I, and the near-IR
followup observations reported in this paper have been obtained relatively close in time (1-2 years),
reducing the effect inaccurate periods will have on the calculations of the phases of the NIR
observations. For most of our objects, the difference between the mean magnitudes derived from the two
independent random-phase JK measurements was comparable to the standard deviations of the individual
observations.

Table 2 gives the intensity mean J and K magnitudes of the individual Cepheids. Each value was
calculated as the average of the individual determinations of the mean magnitude. In Table 2
we also provide the periods (from Paper I) and uncertainties on the mean magnitudes (which contain the
contribution of a 0.03 mag intrinsic error coming from the transformation of the random-phase
to the mean magnitudes; see Soszy{\'n}ski et al. 2005). 

In Figure 2 we show the J- and K-band period-mean magnitude diagrams defined by the NGC 55 Cepheids
in our observed fields. From the 40 stars in these diagrams, we excluded 10 objects from the final distance solution;
these stars are plotted as open circles in Fig. 2, with their IDs indicated. We excluded these
objects from our distance analysis for the reasons given in the following. 
Variables cep001, cep002 and cep004 have extremely long periods
of 176, 152 and 98 days. Cepheids of such long periods, and the corresponding very high luminosities, 
have long been suspected to deviate from the extension of the linear PL relation
defined by Cepheids of shorter periods, in the sense that they are becoming intrinsically
fainter than what the linear PL relation would suggest. The empirical evidence for this effect
has been discussed by a number of authors (e.g. Gieren et al. 2004; Madore \& Freedman 1991).
The position of the three longest-period Cepheids in our sample suggests that the same
effect is present in our data, and we therefore prefer to exclude these objects from the distance
solution. Particularly cep004 at P=98 days is very underluminous in the J band, whereas in K it is closer
to the ridge line. A possible
explanation for this behavior is that this variable suffers excessive reddening. This would be consistent
with its young age, as derived from a period-age relation (e.g. Bono et al. 2005), which implies that
such a long-period Cepheid must still be close to the molecular cloud where it was formed.
An additional reason to exclude Cepheids with periods longer than 100 days in our
solution is the fact that the fiducial LMC PL relation has not been calibrated with such
long-period stars either.
The adopted procedure to introduce an upper period cutoff at 100 days is also consistent with our previous 
Cepheid distance work
in the Araucaria Project. 

Three Cepheids (cep113, cep086 and cep036) are more than one full magnitude brighter in K than
the ridge line luminosities at the corresponding periods, and it seems likely that these stars
are strongly blended by very red objects. This is supported by the observation that these
objects are still over-luminous in the J-band PL relation, but by a smaller amount than in K
(lower panel of Fig. 2). All three objects fall very clearly outside the instability strip on the
K, J-K CMD, which is shown in Fig. 3, supporting the idea that they are either not classical Cepheids,
or stars suffering strong observational anomalies.
The images of these three stars obtained on the nights of best seeing
do indeed suggest the presence of bright companion stars which are not
quite resolved. Yet, it is also possible that these very luminous variables are a different kind of objects,
and not blended Cepheids. They are bright enough for near-IR spectroscopy, which would shed
more light on the true nature of these objects. Whatever the correct explanation will turn out to be
for their excessive brightness, we believe it to be the correct procedure to exclude them
from the distance solution.
 Stars cep079, cep033 and cep013 show very strange PSF profiles which are likely to be caused by
unresolved companion stars as well, consistent with their very bright apparent magnitudes, for their
respective periods. The only Cepheid we have excluded from the sample without having a specific reason other
than its strong deviation, particularly in the K band, from the ridge line is cep023. In K, this
object is 1 mag fainter than the ridge line at this period, with a similar effect in the J band.
Perhaps cep023 is not a classical Cepheid. It could also be an excessively reddened Cepheid.
Below, we show that our distance solution is rather
insensitive to the final choice of the sample we are using for the determination of the
distance of NGC 55. 

In Fig. 4, we plot the PL diagrams in J and K for the adopted final sample of 30 Cepheids. 
Least-squares fits to a line yield slopes of -2.933 $\pm$ 0.133 in K, and -2.843 $\pm$ 0.165
in J, respectively. These slopes are shallower than, but within 2 $\sigma$ consistent with
the slopes of the Cepheid PL relations in the LMC, which are -3.261 in K, and -3.153 in J (Persson et al. 2004).
Following the procedure we have used in our previous papers, we adopt the LMC
slopes of Persson et al. (2004) in our fits. This yields the following PL relations for NGC 55: \\

J = -3.153 log P + (24.395 $\pm$ 0.048)  \hspace*{1cm}    $\sigma$ = 0.265 \\

K = -3.261 log P + (23.975 $\pm$ 0.041)  \hspace*{1cm}    $\sigma$ = 0.223 \\

The zero points in these relations are very little sensitive to our adopted exclusion of suspect objects. If we
retain all 37 Cepheids in the fits except the very strongly and definitively 
blended objects cep113, cep086 and cep036 (which in the
K band are more than a full magnitude brighter than the ridge line luminosity at the respective
periods), and even including the Cepheids with periods longer than 100 days,
the zero points change only slightly to 24.433 in J and 23.996 in K. However, the uncertainties 
of the zero points increase
substantially, and the dispersions of the PL relations
in J and K now increase to 0.405 mag and 0.377 mag, respectively. We are confident that we have cleaned
our raw sample of Cepheids in the best possible way to obtain the most reliable values for the zero points
of the J- and K-band PL relations, and their uncertainties from our datasets. 

In order to determine the relative distance moduli between NGC 55 and the LMC, we need to convert the NICMOS (LCO)
photometric system used by Persson et al. (2004) to the UKIRT system utilized in this paper.
According to Hawarden et al. (2001),
there are just zero point offsets between the UKIRT and NICMOS systems (e.g. no color dependences)
in the J and K filters, which amount to 0.034 and 0.015 mag, respectively. Applying
these offsets, and assuming an LMC true distance modulus of 18.50 as in our previous
work in the Araucaria Project, we derive distance moduli for NGC 55 of 26.593 $\pm$ 0.048 mag
in the J band, and 26.454 $\pm$ 0.041 mag in the K band. 

As in our previous papers in this series,
we adopt the extinction law of Schlegel et al. (1998) and fit a straight line to the relation
$(m-M)_{0} = (m-M)_{\lambda} - A_{\lambda} = (m-M)_{\lambda} - E(B-V) * R_{\lambda}$.
Using the distance moduli in the V and I photometric bands derived in Paper I together with
the values for the J and K bands calculated above, we obtain
for the reddening and the true distance modulus of NGC 55 the following values: \\

$ E(B-V) = 0.127 \pm 0.019$

$(m-M)_{0} = 26.434 \pm 0.037$,

corresponding to a distance of NGC 55 of 1.94 $\pm$ 0.03 Mpc.

In Table 3 we give the adopted values of $R_{\lambda}$ and the unreddened distance moduli
in each band which are obtained with the reddening value determined in our multi-wavelength
approach. The agreement between the de-reddened distance moduli in each band is excellent.
In Fig. 5, we plot the apparent distance moduli in VIJK as a function of $R_{\lambda}$, and the
best fitting straight line to the data; it is appreciated that the total reddening, and the true distance
modulus of NGC 55 are very well determined from this fit.

\section{Discussion}

Our new distance to NGC 55 derived from optical/near-infrared photometry of Cepheids in this galaxy
agrees within the combined 1 $\sigma$ uncertainties with the previous determination from the
Tully-Fisher method (Karachentsev et al. 2003), the I-band magnitude of the tip of the red giant
branch as derived from HST WFPC2/ACS images (Tikhonov et al. 2005), and with the PNLF distance
derived by Van de Steene et al. (2006). Whereas the quoted uncertainty of the TRGB distance is only
slightly larger than the Cepheid distance obtained in this paper, the distances coming from the
TF and PNLF methods are clearly considerably more uncertain. The most discrepant distance
determination from carbon stars in NGC 55 of Pritchet et al. (1987) has probably the largest
systematic uncertainty, due to both the use of a non-reliable distance indicator, and
of an inadequate spatial resolution of 0.59 arcsec/pixel in their CCD images. The short distance
to NGC 55 they derive is likely the result of a strong blending of some of their target carbon
stars by nearby stars not resolved in their photometry. The quopted $\pm$ 0.08 Mpc distance
uncertainty obviously refers to an intrinsic error and does not account for the systematic
uncertainty in this measurement.

An exhaustive discussion of the systematic errors which may affect the Cepheid distances
derived with our multiwavelength technique was presented in the papers of Gieren et al. (2005a, 2006), and
Pietrzy{\'n}ski et al. (2006a) which discussed the multiband Cepheid distance analyses
for NGC 300, NGC 6822 and IC 1613. Here we discuss only the most important and specific
issues concerning NGC 55. First of all, we have been able to use a number of Cepheids
in our present study which is large enough to reduce the effect of a possible inhomogeneous filling of the
Cepheid instability strip to a level that its expected effect on the distance of NGC 55
is insignificant (particularly as our current results from the NIR are combined with
the distance results from the previous optical study in Paper I which were based on more
than 100 Cepheid variables). Also, the removal of outliers is not critical in this study-for all but one
object we have excluded from the sample we adopted for the distance determination we had strong
reasons for the exclusion, and even retaining all objects with the exception of three
blends which are recognized as such from our images produces a change in the distance
which is less than 2\%. 

The near-infrared photometry of this paper has confirmed the distance we had derived
in Paper I from the reddening-independent (V-I) Wesenheit magnitude within $\pm$2\%. Given the
high inclination of NGC 55 with respect to the line of sight and the clear possibility
of relatively strong and variable reddening of the Cepheids in this galaxy, it was imperative
to confirm the results from the Wesenheit index with near-IR photometry. The results of
this paper summarized in Table 3 and Figure 5 suggest that any residual effect of reddening on 
the distance of NGC 55 determined in this paper is negligible. The importance of this cannot
be overstated because in most Cepheid-based distance determinations of late-type galaxies,
particularly when infrared data are sparse or not available, intrinsic reddening is likely
to be the largest source of systematic error. 

Another contributor to the possible systematic error of our present distance result
is the effect of metallicity on the Cepheid PL relation. In the galaxies so far studied in
our project, including NGC 55, we have not found convincing evidence that the slope of the Cepheid PL relation
in optical or near-IR bands is {\it not} universal. The results of Gieren et al. (2005c)
on the Milky Way and LMC Cepheid PL relations from Cepheid distances determined with the
infrared surface brightness technique (Fouqu{\'e} \& Gieren 1997) seem to support the scenario of
a universal {\it slope} of the Cepheid PL relation in optical and NIR bands. This is supported
by the recent parallax work of Benedict et al. (2007), and van Leeuwen et al. (2007),
and also by the very exhaustive recent analysis of Fouqu{\'e} et al. (2007) which all yield
results compatible with a constant slope of the PL relation. However, there is also evidence
for a possible non-constancy of the PL relation slope, as the kink at a period of 10 days
observed in the LMC PL relation in optical bands (e.g. Ngeow \& Kanbur 2006). In the present paper,
we find evidence that the slope of the PL relation in NGC 55 in J and K is the same as in other galaxies,
strengthening the case for the universality of the PL slope,
but this conclusion hinges on the assertion that the objects which are over-bright in K
in Fig. 2 are indeed observational anomalies, as discussed in the previous section.
 Work to settle the question of the universality of the slope of the Cepheid PL relation
  in a definitive way must certainly continue. We intend to give a much more
detailed and quantitative discussion of this issue in a forthcoming paper, and {\it assume}
the constancy of the PL relation slopes in the various bands, as we have done in the previous
papers of this series. 
The extent to which the {\it zero point}
of the PL relation is affected by metallicity is still an open issue and will be addressed
in our project once we will have measured the distances to our target galaxies from other
methods as well, particularly from the blue supergiant FGLR (Kudritzki et al. 2003), and from the
TRGB method. We therefore prefer to leave an exhaustive discussion of this point to a later stage
of our project.

As in previous papers in our project, we have applied utmost care to determine the zero points
of our photometry as accurately as possible. From the arguments given in section 2 we believe that
the zero points are accurate to better than $\pm$0.03 mag, in both J and K. One issue of concern
is the effect of crowding and blending of the target Cepheids in NGC 55 on our results. We were
fortunate enough to have images of exquisite quality, obtained under excellent seeing conditions
at our disposal, and we believe that we were able to identify all Cepheids in our sample whose fluxes are
significantly affected by nearby companion stars. It is, however, difficult to quantify the remaining
effect blending can still have on our distance result as long as we do not have images of our
NGC 55 Cepheids obtained at higher angular resolution. As a guide, however, we can use the results we obtained for
another Sculptor galaxy, NGC 300. For this galaxy, a comparison of single-epoch HST-ACS photometry
with ground-based photometry suggested that the effect
of blending affects the distance determination with our method
by less than 2\% (Bresolin et al. 2005). The effect for NGC 55
should be similar due to its very similar distance, but could be slightly more serious due to
its higher inclination, as compared to NGC 300. We assume that 3\% is a reasonable upper limit
for the possible remaining effect of unresolved stars in the Cepheid photometry on the distance result.
The effects acts to make the Cepheids too bright and therefore tends to decrease the distance.

Probably the largest source of systematic uncertainty on our measured distance to NGC 55 is the value
of the adopted distance to the LMC of 18.50 mag to which our distance determination is tied. The
uncertainty of this value may exceed 10\% (Benedict et al. 2002).
However, if future work changes the currently adopted LMC distance, we can easily
adapt the distances of the target galaxies of our project to the new value. The {\it relative} distance moduli will
remain unaffected.

From this discussion, and from the conclusions about systematic uncertainties presented in the
previous papers in this series we conclude that, apart from the systematic uncertainty on the
adopted distance to the LMC, the total systematic error on our present distance
determination of NGC 55 does not exceed $\sim$4\%. We believe that our combined optical-NIR 
Cepheid work on NGC 55
has brought about a distance determination to this galaxy which is clearly more accurate than
the previous attempts to measure the distance to NGC 55 listed in Table 4.

\section{Summary and Conclusions}
We have carried out the first Cepheid-based distance determination for the Sculptor Group spiral
galaxy NGC 55 from deep images in the optical/near-infrared VIJK bands using our multiwavelength
approach used in earlier papers of this series. The resulting distance has a random uncertainty
of about 2\% and an estimated systematic uncertainty of 4\%, not taking account the
uncertainty of the LMC distance to which our NGC 55 distance result is tied. Our distance determination
is virtually unaffected by the reddening of the NGC 55 Cepheids which we have determined very
accurately in our procedure. 

In spite of our effort to recognize Cepheids in our database showing
clear signs of being blended with nearby stars in our images, and exclude such stars in the distance analysis,
it is likely that blending of the remaining Cepheids with non-resolved companion stars
is the single most important source of systematic uncertainty in the present study, with an estimated
$\sim$3\% effect, with this estimation coming from our former HST-based work on NGC 300, where the blending effect on the
Cepheid distance was found to be less than 2\%. The blending effect in the case of NGC 55 is likely to be 
somewhat more severe due to the larger inclination of the galaxy with respect to the line of sight.

Our Cepheid distance to NGC 55 is more accurate than the existing distance estimates from
the PNLF, TRGB and Tully-Fisher methods, and constitutes another step in our effort to improve the determination 
of the metallicity dependence of stellar methods of distance measurement, including Cepheids, from
comparative analyses of the distances of a set of Local Group and Sculptor Group galaxies in our
Araucaria Project. Such studies will be the subject of forthcoming papers. 

Within the (small) uncertainties, the Cepheid distances of the Sculptor galaxies NGC 55 and NGC 300 agree.
Taking into account the small angular separation of these two galaxies in the sky, the distance
of NGC 55 measured in this paper supports the conclusion that both galaxies are physically
associated.

\acknowledgments
WG, GP and DM gratefully acknowledge financial support for this
work from the Chilean Center for Astrophysics FONDAP 15010003. 
Support from the Polish grant N203 002 31/046 is also acknowledged.
IS was supported by the Foundation for Polish Science through the Homing Programme. 
It is a special pleasure to thank the support astronomers at ESO-Paranal
for their expert help in the observations, and the ESO OPC for the
generous amounts of observing time at VLT allocated to our Large Programme.
We thank an anonymous referee for his comments.

\begin{figure}[p] 
\vspace*{18cm}
\includegraphics{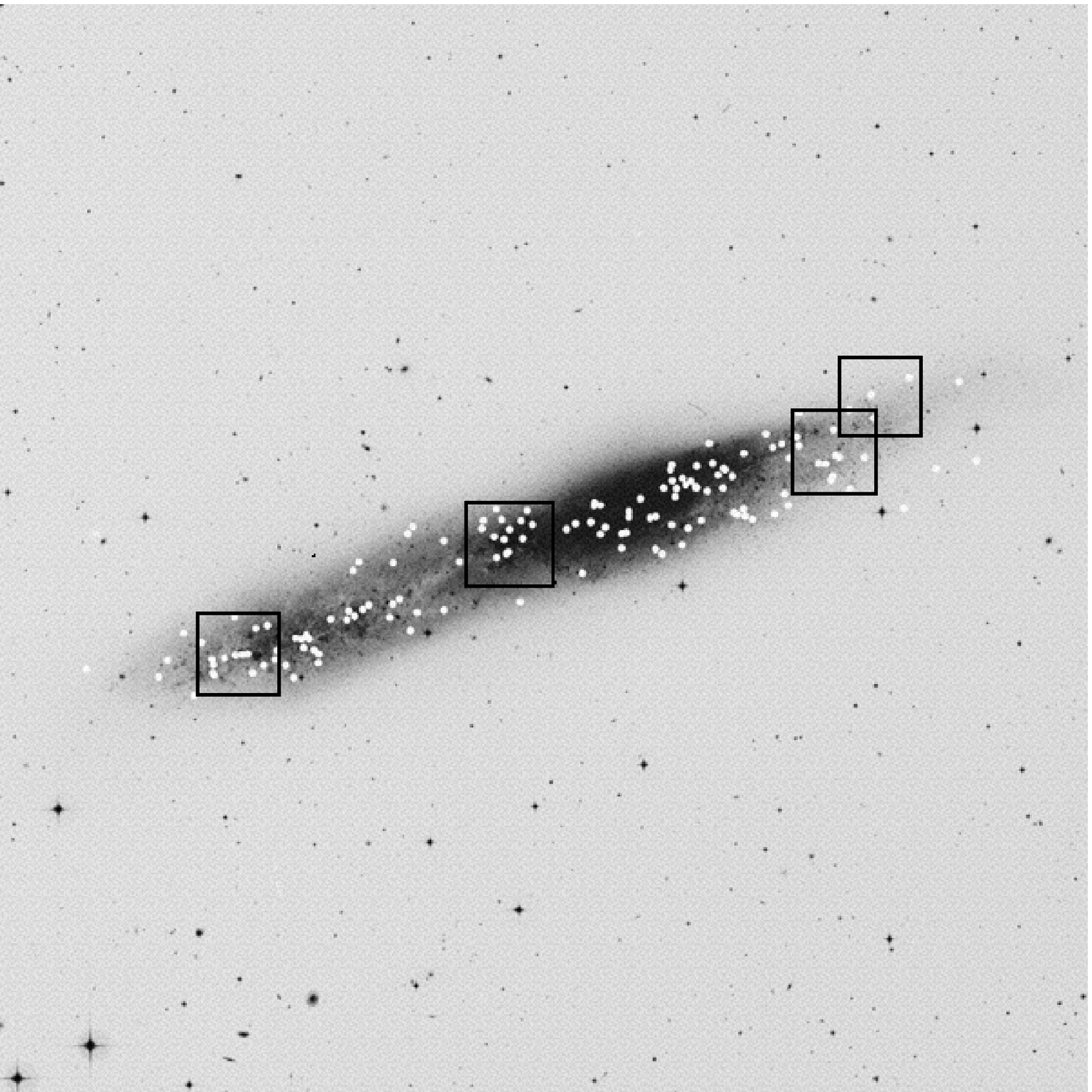} 
\caption{The location of the four observed VLT/ISAAC fields in NGC 55 on the DSS
blue plate. We observed each field on two different nights. The white dots indicate
the Cepheids discovered in our previous optical survey (Paper I).}
\end{figure}

\begin{figure}[htb]
\vspace*{20cm}
\includegraphics{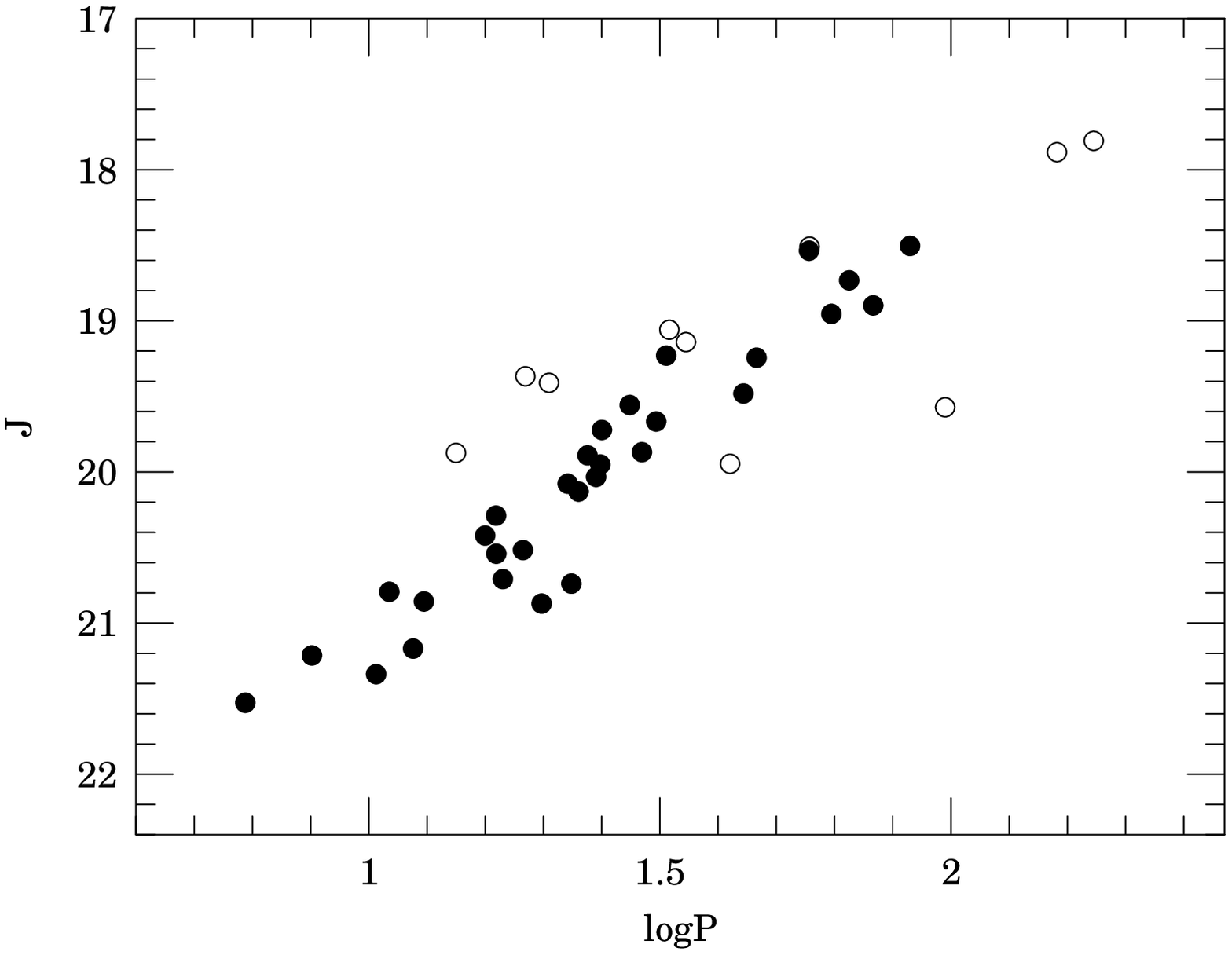}
\includegraphics{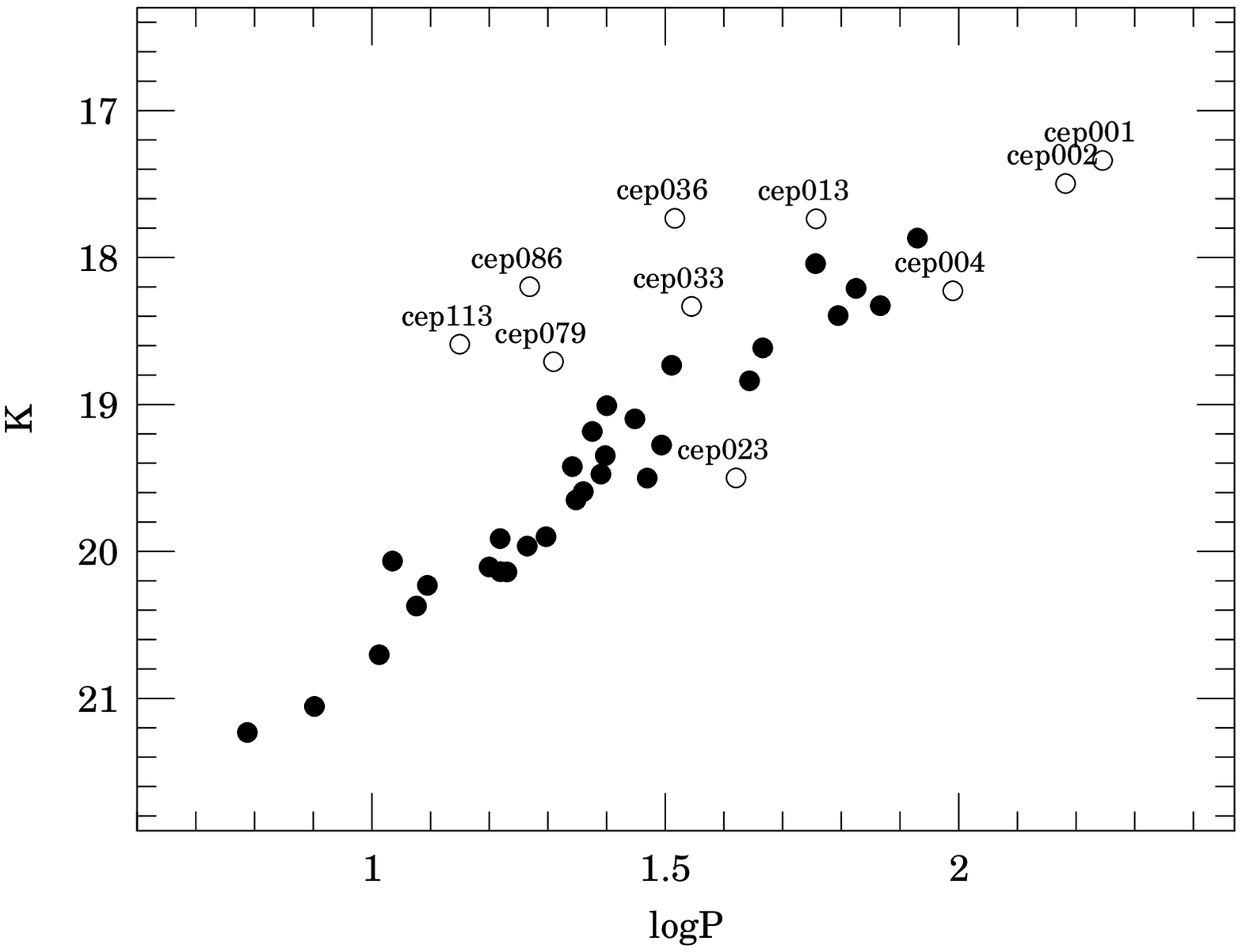}
\caption{The near-infrared J and K band period-luminosity relations
defined by the 40 observed NGC 55 Cepheids.
Each mean magnitude was derived from two random-phase
observations by the method of Soszy{\'n}ski et al. (2005). Stars which
were not used in the
distance determination (see text) are indicated by open circles.}
\end{figure}

\begin{figure}[htb]
\vspace*{17cm}
\includegraphics{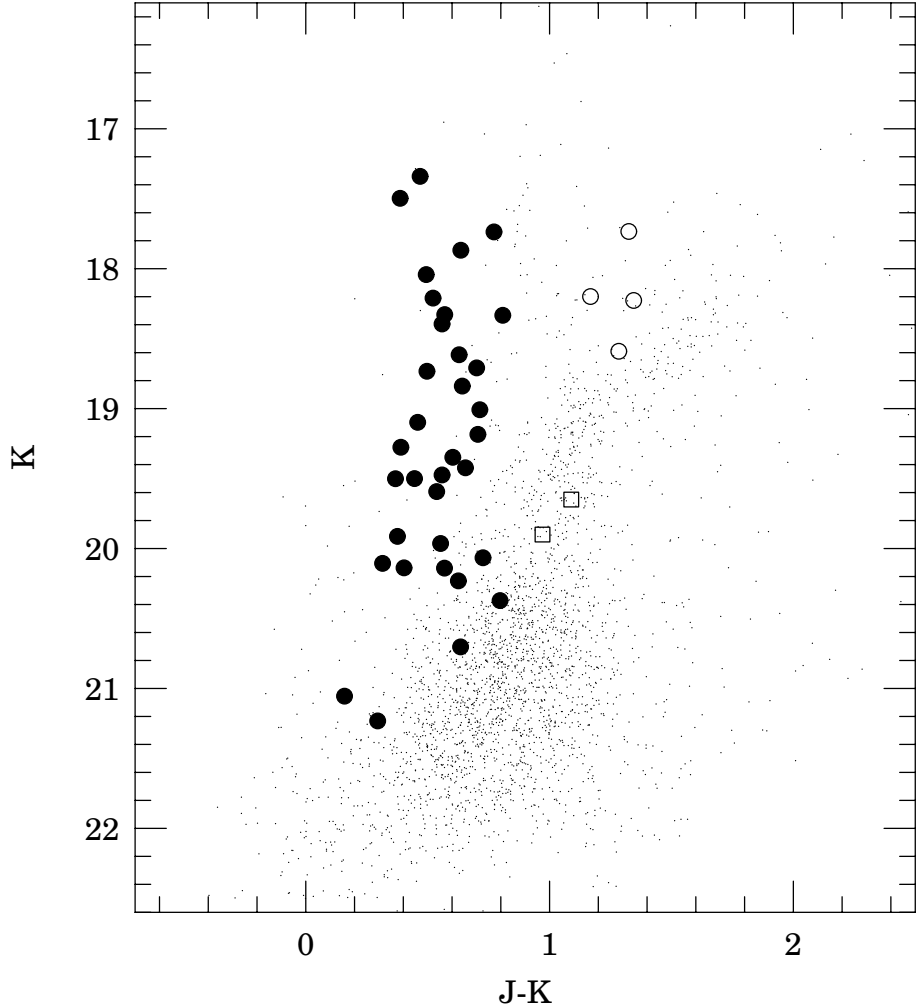}
\caption{The location of the Cepheid variables in NGC 55 in the K, J-K
color-magnitude diagram. The bulk of the variables clearly define
the instability strip in its expected position (filled circles).
Outliers are a group
of four very bright and red stars (open circles), which are objects
cep004, 036,
086 and 113 in our database. Cepheids 036, 086 and 113 are the most
deviating objects in the K-band PL diagram in Fig. 2, supporting
our choice to remove these objects in the distance solution. These
objects are either heavily blended by very red and bright objects, or
they are not classical Cepheids. There are two additional objects
located outside the instability strip (open squares), cep69 and cep81.
If all six outliers are removed from the database, the remaining
Cepheids lying inside the instability strip define PL relation slopes
of -2.81 +/- 0.17 in J, and -2.94 +/- 0.20 in K, which are consistent
with the LMC PL relation slopes of Persson et al. within 2 sigma.}
\end{figure}

\begin{figure}[htb]
\vspace*{20cm}
\includegraphics{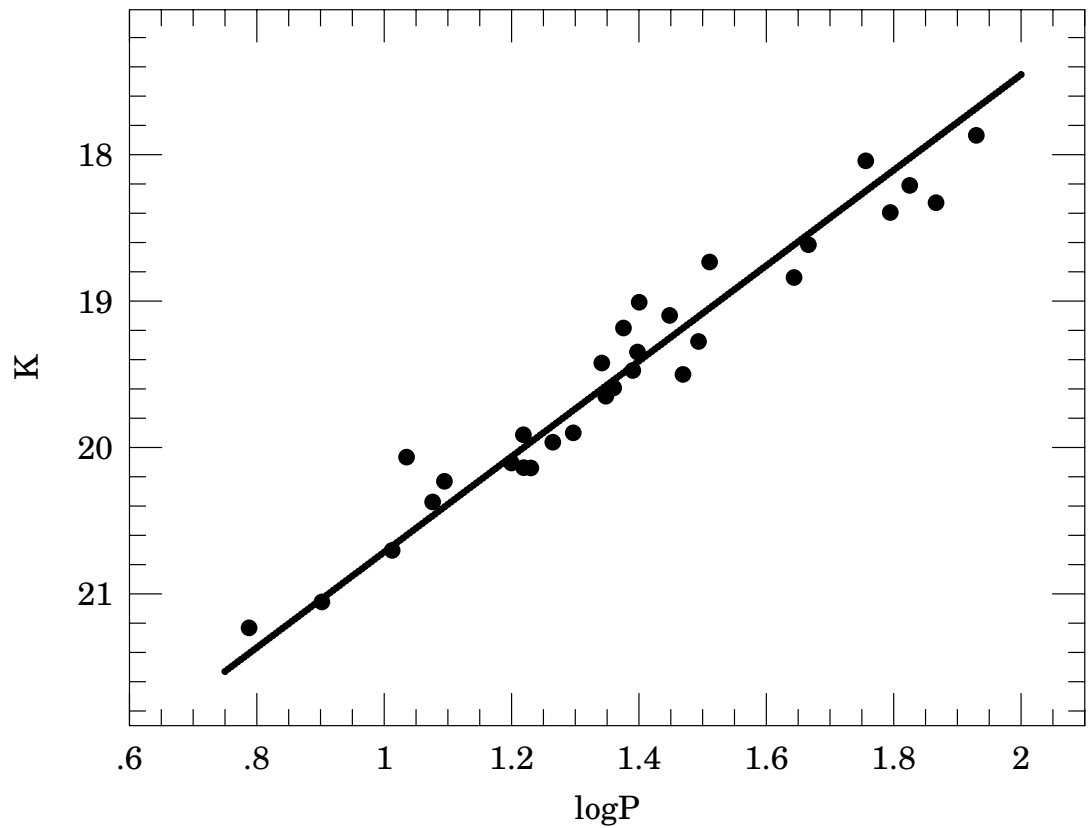}
\includegraphics{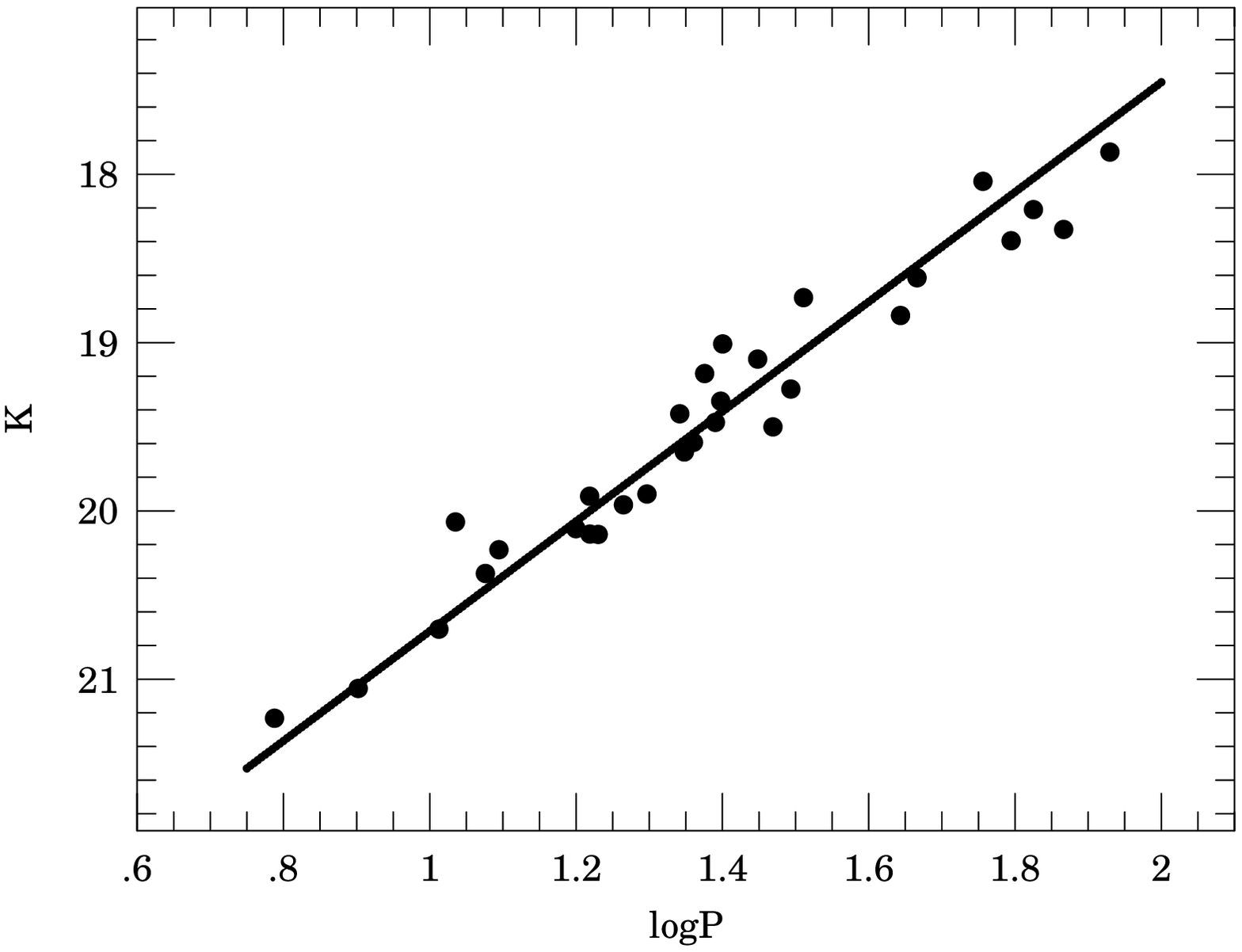}
\caption{Cepheids in NGC 55 adopted for the near-infrared PL solutions, plotted along with the best-fitting lines.
The slopes of the fits were adopted from the LMC, and the zero points determined
from our data. }
\end{figure}

\begin{figure}[p]
\includegraphics{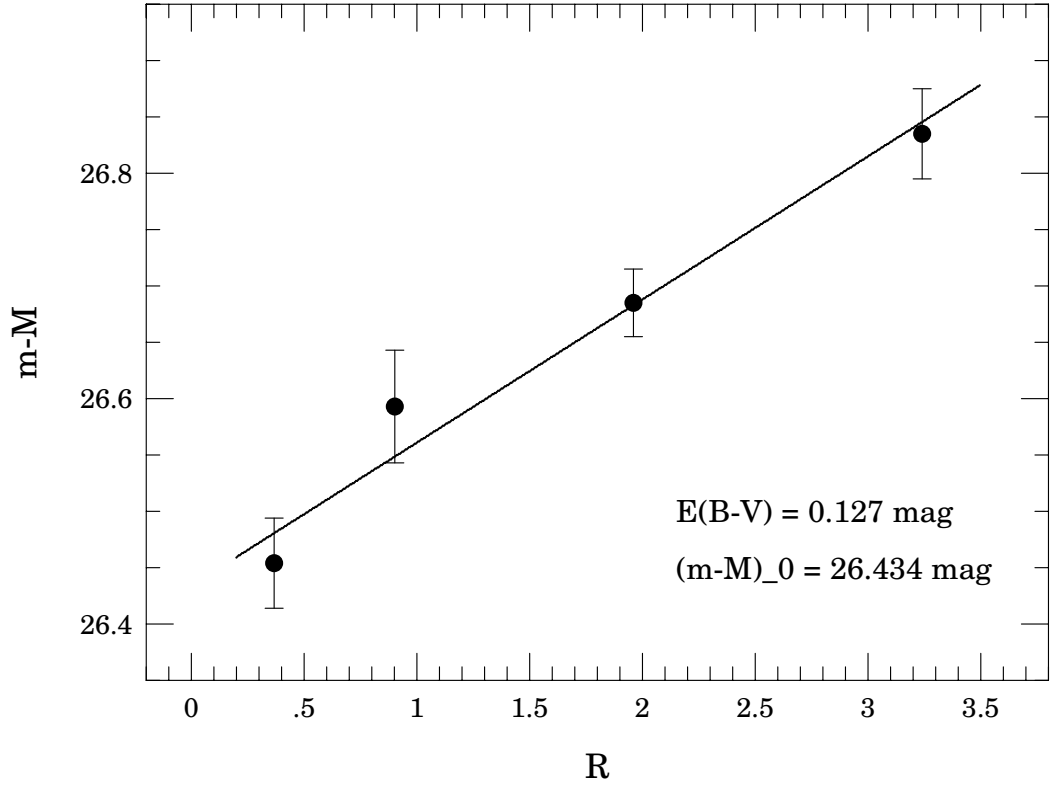}
\vspace{10cm}
\caption{Apparent distance moduli to NGC 55 as derived in the VIJK photometric bands,
plotted against the ratio of total to selective extinction as adopted from
the Schlegel et al. reddening law. The intersection and
slope of the best-fitting line give the true distance modulus and the average total
reddening, respectively. The data in this diagram suggest that the Galactic reddening law
is a very good approximation for NGC 55 as well.}
\end{figure}

\clearpage
\begin{deluxetable}{c c c c c c c}
\tablewidth{0pc}
\tablecaption{Journal of the Individual J and K band Observations of NGC 55
Cepheids}
\tablehead{ \colhead{ID} & \colhead{J HJD} & \colhead{J}  & \colhead{$\sigma$} 
& \colhead{K HJD} & \colhead{K} & \colhead{$\sigma$}\\
\colhead{} & \colhead{2400000+} & \colhead{mag} & \colhead{mag} & \colhead{2400000+} &
\colhead{mag} & \colhead{mag}
 }
\startdata
cep001 &  53206.335935 &   17.751 &   0.009 &  53206.260866 &  17.285 &   0.013 \\ 
cep010 &  53206.335935 &   18.786 &   0.014 &  53206.260866 &  18.299 &   0.017 \\ 
cep013 &  53206.335935 &   18.602 &   0.013 &  53206.260866 &  17.798 &   0.015 \\ 
cep126 &  53206.335935 &   20.992 &   0.061 &  53206.260866 &  20.132 &   0.072 \\ 
cep002 &  53222.266831 &   17.820 &   0.024 &  53222.299022 &  17.435 &   0.025 \\ 
cep004 &  53222.266831 &   19.682 &   0.028 &  53222.299022 &  18.259 &   0.039 \\ 
cep005 &  53222.266831 &   18.389 &   0.024 &  53222.299022 &  17.781 &   0.023 \\ 
cep009 &  53222.266831 &   18.972 &   0.021 &  53222.299022 &  18.403 &   0.023 \\ 
cep012 &  53222.266831 &   18.739 &   0.025 &  53222.299022 &  18.232 &   0.021 \\ 
cep036 &  53222.266831 &   19.170 &   0.029 &  53222.299022 &  17.779 &   0.044 \\ 
cep055 &  53222.266831 &   19.719 &   0.031 &  53222.299022 &  18.957 &   0.032 \\ 
cep072 &  53222.266831 &   19.911 &   0.039 &  53222.299022 &  19.336 &   0.029 \\ 
cep081 &  53222.266831 &   21.152 &   0.058 &  53222.299022 &  19.914 &   0.060 \\ 
cep002 &  53249.278595 &   17.882 &   0.022 &  53249.200628 &  17.515 &   0.026 \\ 
cep004 &  53249.278595 &   19.631 &   0.037 &  53249.200628 &  18.298 &   0.029 \\ 
cep005 &  53249.278595 &   18.634 &   0.023 &  53249.200628 &  17.918 &   0.023 \\ 
cep009 &  53249.278595 &   18.918 &   0.022 &  53249.200628 &  18.335 &   0.028 \\ 
cep012 &  53249.278595 &   19.213 &   0.028 &  53249.200628 &  18.536 &   0.032 \\ 
cep036 &  53249.278595 &   19.126 &   0.035 &  53249.200628 &  17.854 &   0.027 \\ 
cep055 &  53249.278595 &   19.688 &   0.036 &  53249.200628 &  18.929 &   0.037 \\ 
cep072 &  53249.278595 &   20.117 &   0.039 &  53249.200628 &  19.308 &   0.044 \\ 
cep081 &  53249.278595 &   20.734 &   0.079 &  53249.200628 &  20.047 &   0.120 \\ 
cep020 &  53204.306992 &   19.404 &   0.026 &  53204.228037 &  18.894 &   0.023 \\ 
cep023 &  53204.306992 &   19.924 &   0.032 &  53204.228037 &  19.510 &   0.022 \\ 
cep037 &  53204.306992 &   19.149 &   0.020 &  53204.228037 &  18.603 &   0.016 \\ 
cep039 &  53204.306992 &   19.578 &   0.040 &  53204.228037 &  19.259 &   0.026 \\ 
cep079 &  53204.306992 &   19.308 &   0.035 &  53204.228037 &  18.469 &   0.015 \\ 
cep087 &  53204.306992 &   20.296 &   0.035 &  53204.228037 &  19.862 &   0.027 \\ 
cep098 &  53204.306992 &   20.878 &   0.046 &  53204.228037 &  20.016 &   0.045 \\ 
cep100 &  53204.306992 &   20.691 &   0.057 &  53204.228037 &  20.306 &   0.049 \\ 
cep101 &  53204.306992 &   20.168 &   0.048 &  53204.228037 &  19.850 &   0.030 \\ 
cep138 &  53204.306992 &   21.015 &   0.095 &  53204.228037 &  20.828 &   0.067 \\ 
cep020 &  53223.250368 &   19.430 &   0.020 &  53223.275357 &  18.702 &   0.027 \\ 
cep023 &  53223.250368 &   19.909 &   0.023 &  53223.275357 &  19.456 &   0.034 \\ 
cep037 &  53223.250368 &   19.267 &   0.014 &  53223.275357 &  18.850 &   0.024 \\ 
cep039 &  53223.250368 &   19.796 &   0.027 &  53223.275357 &  19.354 &   0.033 \\ 
cep069 &  53223.250368 &   20.831 &   0.075 &  53223.275357 &  19.645 &   0.068 \\ 
cep079 &  53223.250368 &   19.321 &   0.022 &  53223.275357 &  18.745 &   0.026 \\ 
cep087 &  53223.250368 &   20.298 &   0.024 &  53223.275357 &  19.873 &   0.037 \\ 
cep098 &  53223.250368 &   20.865 &   0.038 &  53223.275357 &  20.482 &   0.105 \\ 
cep100 &  53223.250368 &   20.514 &   0.028 &  53223.275357 &  20.037 &   0.049 \\ 
cep101 &  53223.250368 &   20.193 &   0.029 &  53223.275357 &  19.778 &   0.036 \\ 
cep126 &  53223.250368 &   21.012 &   0.073 &  53223.275357 &  20.528 &   0.126 \\ 
cep138 &  53223.250368 &   21.098 &   0.055 &  53223.275357 &  20.959 &   0.095 \\ 
cep014 &  53246.212176 &   18.745 &   0.028 &  53246.131676 &  18.213 &   0.016 \\ 
cep018 &  53246.212176 &   19.072 &   0.031 &  53246.131676 &  18.439 &   0.023 \\ 
cep033 &  53246.212176 &   19.011 &   0.027 &  53246.131676 &  18.299 &   0.019 \\ 
cep043 &  53246.212176 &   19.889 &   0.057 &  53246.131676 &  19.571 &   0.025 \\ 
cep059 &  53246.212176 &   19.656 &   0.038 &  53246.131676 &  19.080 &   0.021 \\ 
cep044 &  53246.212176 &   19.747 &   0.044 &  53246.131676 &  19.172 &   0.036 \\ 
cep056 &  53246.212176 &   19.753 &   0.076 &  53246.131676 &  19.770 &   0.022 \\ 
cep057 &  53246.212176 &   19.801 &   0.036 &  53246.131676 &  19.081 &   0.030 \\ 
cep064 &  53246.212176 &   20.031 &   0.038 &  53246.131676 &  19.520 &   0.027 \\ 
cep086 &  53246.212176 &   19.554 &   0.019 &  53246.131676 &  18.289 &   0.021 \\ 
cep103 &  53246.212176 &   20.404 &   0.193 &  53246.131676 &  21.037 &   0.037 \\ 
cep113 &  53246.212176 &   19.780 &   0.048 &  53246.131676 &  18.585 &   0.060 \\ 
cep122 &  53246.212176 &   20.896 &   0.067 &  53246.131676 &  20.251 &   0.045 \\ 
cep129 &  53246.212176 &   99.999 &   0.118 &  53246.131676 &  20.168 &   9.999 \\ 
cep014 &  53270.216781 &   18.395 &   0.020 &  53270.125592 &  17.930 &   0.024 \\ 
cep018 &  53270.216781 &   19.435 &   0.027 &  53270.125592 &  18.793 &   0.027 \\ 
cep033 &  53270.216781 &   18.991 &   0.031 &  53270.125592 &  18.086 &   0.045 \\ 
cep043 &  53270.216781 &   19.788 &   0.024 &  53270.125592 &  19.240 &   0.028 \\ 
cep059 &  53270.216781 &   19.811 &   0.032 &  53270.125592 &  19.161 &   0.045 \\ 
cep044 &  53270.216781 &   19.723 &   0.029 &  53270.125592 &  19.277 &   0.037 \\ 
cep056 &  53270.216781 &   19.876 &   0.022 &  53270.125592 &  19.287 &   0.029 \\ 
cep057 &  53270.216781 &   19.892 &   0.035 &  53270.125592 &  19.500 &   0.037 \\ 
cep064 &  53270.216781 &   19.913 &   0.028 &  53270.125592 &  19.452 &   0.031 \\ 
cep086 &  53270.216781 &   19.251 &   0.017 &  53270.125592 &  18.232 &   0.024 \\ 
cep103 &  53270.216781 &   20.504 &   0.048 &  53270.125592 &  20.250 &   0.062 \\ 
cep113 &  53270.216781 &   20.134 &   0.045 &  53270.125592 &  18.482 &   0.029 \\ 
cep122 &  53270.216781 &   20.759 &   0.038 &  53270.125592 &  20.073 &   0.054 \\ 
cep129 &  53270.216781 &   20.649 &   0.040 &  53270.125592 &  20.057 &   0.072 \\ 
cep132 &  53270.216781 &   21.590 &   0.094 &  53270.125592 &  20.881 &   0.096 \\ 
cep142 &  53270.216781 &   21.655 &   0.127 &  53270.125592 &  21.279 &   0.145 \\ 
\enddata
\end{deluxetable}

\clearpage

\begin{deluxetable}{c r c c c c}
\tablecaption{Intensity mean J and K magnitudes for 40 Cepheid variables in NGC 55}
\tablehead{
\colhead{ID} & \colhead{P} &
\colhead{$<J>$} & \colhead{$\sigma_{\rm J}$} & \colhead{$<K>$} &
\colhead{$\sigma_{\rm K}$}\\ 
& \colhead{days} & \colhead{mag} & \colhead{mag} & \colhead{mag} &
\colhead{mag}
}
\startdata
cep001 & 175.9086 &  17.809 &   0.027 &  17.340 &   0.028 \\ 
cep002 & 152.0943 &  17.884 &   0.034 &  17.497 &   0.036 \\ 
cep004 &  97.7291 &  19.572 &   0.041 &  18.227 &   0.042 \\ 
cep005 &  85.0550 &  18.504 &   0.034 &  17.868 &   0.034 \\ 
cep009 &  73.5323 &  18.898 &   0.033 &  18.328 &   0.036 \\ 
cep010 &  66.8528 &  18.732 &   0.029 &  18.210 &   0.030 \\ 
cep012 &  62.3186 &  18.954 &   0.036 &  18.395 &   0.037 \\ 
cep013 &  57.1544 &  18.509 &   0.028 &  17.737 &   0.029 \\ 
cep014 &  57.0377 &  18.536 &   0.035 &  18.042 &   0.032 \\ 
cep018 &  46.3493 &  19.244 &   0.038 &  18.615 &   0.035 \\ 
cep020 &  44.0077 &  19.481 &   0.034 &  18.839 &   0.035 \\ 
cep023 &  41.7395 &  19.946 &   0.037 &  19.500 &   0.038 \\ 
cep033 &  35.0516 &  19.141 &   0.037 &  18.333 &   0.031 \\ 
cep036 &  32.8210 &  19.059 &   0.041 &  17.734 &   0.044 \\ 
cep037 &  32.4267 &  19.230 &   0.030 &  18.733 &   0.032 \\ 
cep039 &  31.1574 &  19.666 &   0.042 &  19.276 &   0.039 \\ 
cep043 &  29.4512 &  19.869 &   0.050 &  19.501 &   0.036 \\ 
cep044 &  28.0657 &  19.557 &   0.045 &  19.098 &   0.044 \\ 
cep055 &  25.1401 &  19.722 &   0.042 &  19.008 &   0.043 \\ 
cep056 &  24.9857 &  19.951 &   0.033 &  19.348 &   0.036 \\ 
cep057 &  24.5666 &  20.033 &   0.043 &  19.474 &   0.042 \\ 
cep059 &  23.7508 &  19.890 &   0.043 &  19.184 &   0.043 \\ 
cep064 &  22.9247 &  20.130 &   0.042 &  19.593 &   0.038 \\ 
cep069 &  22.2824 &  20.739 &   0.079 &  19.650 &   0.072 \\ 
cep072 &  21.9627 &  20.078 &   0.046 &  19.423 &   0.045 \\ 
cep079 &  20.3942 &  19.410 &   0.038 &  18.709 &   0.033 \\ 
cep081 &  19.8080 &  20.871 &   0.074 &  19.900 &   0.098 \\ 
cep086 &  18.5752 &  19.367 &   0.031 &  18.199 &   0.034 \\ 
cep087 &  18.3966 &  20.517 &   0.039 &  19.964 &   0.041 \\ 
cep098 &  16.9912 &  20.709 &   0.049 &  20.140 &   0.085 \\ 
cep100 &  16.5556 &  20.541 &   0.051 &  20.138 &   0.055 \\ 
cep101 &  16.5403 &  20.289 &   0.047 &  19.913 &   0.042 \\ 
cep103 &  15.8413 &  20.421 &   0.054 &  20.106 &   0.057 \\ 
cep113 &  14.1150 &  19.874 &   0.052 &  18.590 &   0.053 \\ 
cep122 &  12.4326 &  20.857 &   0.060 &  20.231 &   0.056 \\ 
cep126 &  11.9130 &  21.169 &   0.074 &  20.372 &   0.076 \\ 
cep129 &  10.8421 &  20.793 &   0.047 &  20.066 &   0.076 \\ 
cep132 &  10.2913 &  21.338 &   0.097 &  20.703 &   0.099 \\ 
cep138 &   7.9824 &  21.214 &   0.060 &  21.055 &   0.098 \\ 
cep142 &   6.1346 &  21.527 &   0.129 &  21.232 &   0.147 \\ 
\enddata
\end{deluxetable}

\begin{deluxetable}{cccccc}
\tablewidth{0pc}
\tablecaption{Reddened and Absorption-Corrected Distance Moduli for NGC
55 in Optical and Near-Infrared Bands}
\tablehead{ \colhead{Band} & $V$ & $I$ & $J$ & $K$ & $E(B-V)$ }
\startdata
$m-M$                &   26.835 &  26.685 &  26.559 &  26.439 &   --  \nl
${\rm R}_{\lambda}$  &   3.24   &  1.96   &  0.902  &  0.367  &   --  \nl
$(m-M)_{0}$           &   26.423 &  26.436 &  26.444 &  26.392 &  0.127 \nl
\enddata
\end{deluxetable}

\begin{deluxetable}{lll}
\tablewidth{0pc}
\tablecaption{Table 4.  Previous Distance Determinations to NGC 55}
\tablehead{ Distance [Mpc] & Method & Reference} 
\startdata
  1.34 +/- 0.08 &     Carbon stars &        Pritchet et al. 1987 \nl
  1.8  +/- 0.2  &     Tully-Fisher &       Karachentsev et al. 2003 \nl
  2.12 +/- 0.10 &     TRGB         &       Tikhonov et al. 2005 \nl
  2.30 +/- 0.35 &     PNLF         &       Van de Steene et al. 2006 \nl
  1.91 +/- 0.10 &     Cepheids, VI &       Pietrzynski et al. 2006b \nl
  1.94 +/- 0.08 &     Cepheids, VIJK &     this paper \nl
\enddata
\end{deluxetable}

\end{document}